\documentclass[12pt,a4paper]{article}

\usepackage{amsmath}
\usepackage{amsfonts}

\numberwithin{equation}{section}

\renewcommand{\title}[1]{{\center\large\textbf{#1}}}

\begin{document}


\begin{flushright}
  hep-th/0109052
\end{flushright}
\vfill
\title{Dynamics of the spinning particle in one dimension}
\begin{center}
  W. Machin\footnote{email: \texttt{wmachin@mth.kcl.ac.uk}}\\
  \textit{King's College London, Strand, London WC2R 1LS}\\
\end{center}
\begin{abstract}
  The most general $N=1$ Lagrangian for the spinning particle with
  local supersymmetry is found and the constraints of the system are
  analysed. The Dirac quantisation of the model is also investigated.
\end{abstract}
\vfill
\newpage
\section{Introduction}


The classical and quantum properties of a particle propagating in a
space have found many applications from general relativity to quantum
mechanics and quantum field theory. In the past the model that was
mostly investigated was a relativistic or non-relativistic particle
propagating in a Riemannian manifold with a metric $g$. These results
were extended to $N=1$ supersymmetric particle
models in~\cite{Brink:1977uf}. Later the action of supersymmetric
particles with extended supersymmetry was given in~\cite{Howe:1988ft}.

In~\cite{Coles:1990hr} it was found that rigid supersymmetry in one
dimension allows for the construction of more general models than
those considered in the past.  In particular, the manifold that the
particle propagates in can have torsion which is not a closed 3-form.
Such models have been found to describe the effective theory of
multi-black hole systems~\cite{Maloney:1999dv}. A scalar potential and
a coupling to a magnetic field in the action of~\cite{Coles:1990hr} was added
in~\cite{Hull:1993ct}.

In particular it was found in~\cite{Coles:1990hr} that the $N=1$
particle with rigid supersymmetry is described by two multiplets
$q^{\mu}$ and $\chi^{\alpha}$. The superfield $q$ has components
$q^{\mu}$ which are the positions of the particle and a worldline
fermion $\lambda^{\mu}$.  The components of $\chi^{\alpha}$ are a
worldline fermion $\chi^{\alpha}$ and an auxiliary field $y^{\alpha}$.

In this paper we give the action of an $N=1$ supersymmetric
relativistic particle propagating in a manifold and with couplings
which include a metric $g$, torsion $c$, an electromagnetic gauge
potential $A$ and a scalar potential $V$. For this we introduce an
einbein $e$ and a worldline gravitino $\psi$. Then we give the
transformation law of these fields under local supersymmetry and show
that our action is invariant up to surface terms. The construction of
such an $N=1$ locally supersymmetric action has been considered
before~\cite{vanHolten:1995qt}. However in our formalism the geometric
interpretation of the various couplings is manifest in the manner
of the rigid models in~\cite{Coles:1990hr}. In addition we consider
more general couplings.

Furthermore we use Dirac's method to analyse the constraints of the
system and compute the Poisson bracket algebra of constraints, and
then discuss the Dirac quantisation of the model. As expected the
constraint associated to supersymmetry is a Dirac operator but now it
involves the torsion and it is twisted with respect to a gauge field.
The square of the supersymmetry constraint gives the Hamiltonian
constraint.

The paper is organised as follows. In section~\ref{sec:review}, the
sigma models with rigid supersymmetry are reviewed. In
section~\ref{sec:local}, the $N=1$ action and local supersymmetries
are given. The constraints of the system are analysed in
section~\ref{sec:constraints} and in section~\ref{sec:quantise} its
Dirac quantisation is investigated.





\section{The $N=1$ supersymmetric action}
\label{sec:review}
The spinning particle model we use is a one-dimensional sigma model
with a worldline superspace $\Xi$ and target manifold $\mathcal{M}$.
$\Xi$ is parameterised by one real coordinate $t$, representing time,
and one real anticommuting coordinate $\theta$.

The real bosonic superfield $q^{\mu}(t,\theta)$ describes the position
of the spinning particle and its classical spin degrees of freedom. It
is a map from the base space into the target manifold in which the
particle lives,
\begin{equation}
  q: \Xi \rightarrow \mathcal{M}
\end{equation}

There is also a fermionic superfield $\chi^{\alpha}(t,\theta)$ which
can be thought of as a section of the bundle $q^*\epsilon$ over $\Xi$
where $\epsilon$ is a vector bundle over
$\mathcal{M}$~\cite{Coles:1990hr}. It represents the Yang-Mills sector
of the theory and is necessary for the inclusion of a potential term,
and generalises the axial spinning coordinate $\psi_5$
in~\cite{Brink:1977uf}.

The most general $N=1$ spinning particle action with rigid
supersymmetry is~\cite{Papadopoulos:2000ka},
\begin{equation} 
  \label{eq:rigidactionsf}
  \begin{split}
    S=-\int\mspace{-5mu} \mathrm{d}t\,\mathrm{d}\theta 
    \biggl[ 
    &\frac{1}{2} g_{\mu\nu} Dq^{\mu}\partial_t q^{\nu} 
    - \frac{1}{2} h_{\alpha\beta} \chi^{\alpha} \nabla \chi^{\beta}
    + \frac{1}{3!} c_{\mu\nu\rho} Dq^{\mu} Dq^{\nu} Dq^{\rho} \\
    &+ \frac{1}{2} m_{\mu\alpha\beta} Dq^{\mu} \chi^{\alpha} \chi^{\beta}             
    + \frac{1}{2} n_{\mu\nu\alpha}Dq^{\mu}Dq^{\nu}\chi^{\alpha}
    + \frac{1}{3!} l_{\alpha\beta\gamma}\chi^{\alpha}\chi^{\beta}\chi^{\gamma} \\
    &- f_{\mu\alpha}\partial_t q^{\mu}\chi^{\alpha} 
    + A_{\mu}Dq^{\mu}
    +ms_{\alpha}\chi^{\alpha}
    \biggr]
  \end{split}
\end{equation}
The superderivative $D$ is defined by
$D=\partial_\theta-\theta\partial_t$ and $\nabla$ is its covariant
version with respect to the connection on the fibre bundle,
\begin{equation}
  \nabla\chi^{\alpha} = D\chi^{\alpha} + Dq^{\mu} B_{\mu\phantom{\alpha}\beta}^{\phantom{\mu}\alpha} \chi^{\beta}
\end{equation}
where $B_{\mu\phantom{\alpha}\beta}^{\phantom{\mu}\alpha}$ is a
connection of the bundle $\epsilon$ with fibre metric
$h_{\alpha\beta}$. We may assume that the fibre metric
$h_{\alpha\beta}$ is compatible with the connection,
$\nabla_{\mu}h_{\alpha\beta}=0$.

The action includes the torsion term $c_{\mu\nu\rho}$ and an
electromagnetic potential $A_{\mu}$. The potential of the theory is
described in terms of $s_{\alpha}$~\cite{Hull:1993ct},
$V(q)=ms_{\alpha}s^{\alpha}/2$. Also present are the Yukawa couplings
$m_{\mu\alpha\beta},n_{\mu\nu\alpha},l_{\alpha\beta\gamma}$ and
$f_{\mu\alpha}$.


The action is invariant under worldline translations generated by
$H=\partial_t$,
\begin{align}
  \delta^{(H)}_{\epsilon} q^{\mu}&=\epsilon\Dot{q}^{\mu} 
& \delta^{(H)}_{\epsilon} \chi^{\mu}&=\epsilon\Dot{\chi}^{\mu} 
\end{align}
and supersymmetry transformations generated by $Q=\partial_\theta+\theta\partial_t$,
\begin{align}
\label{eq:rst}
  \delta_\zeta q^{\mu} &= \zeta Q q^{\mu} 
  & \delta_\zeta \chi^{\alpha} &= \zeta Q \chi^{\alpha}
\end{align}
where $\zeta$ is anticommuting. The algebra of these two transformations is
\begin{equation}
  [ \delta_\zeta, \delta_\eta ] q^{\mu} = 2 \zeta\eta \Dot{q}^{\mu}
 = 2\delta^{(H)}_{\zeta\eta} q^{\mu}
\end{equation}
which realises the commutator $\{Q,Q\}=2H$ of the $N=1$ supersymmetry
algebra. 

We expand the action in components defined by
\begin{align}
  q^{\mu} &= q^{\mu}| & \chi^{\alpha} &= \chi^{\alpha}| \\
  \lambda^{\mu} &= Dq^{\mu}| & y^{\alpha} &= \nabla \chi^{\alpha}|
\end{align}
where the line means evaluation at $\theta=0$. We are following the standard
convention of using the same letter for a superfield and its lowest
component. The field $y^{\alpha}$ is auxiliary and we will eliminate it later.
Expanding the action~(\ref{eq:rigidactionsf}) into component form, we
find that
\begin{equation}
  \label{eq:rigidactioncom}
  \begin{split}
    S=\int\mspace{-5mu} \mathrm{d}t
    \biggl[ 
    & \frac{1}{2} g_{\mu\nu} \Dot{q}^{\mu}\Dot{q}^{\nu}
    + \frac{1}{2} g_{\mu\nu} \lambda^{\mu} \nabla^{(+)}_t \lambda^{\nu} 
    - \frac{1}{2}h_{\alpha\beta} y^{\alpha}y^{\beta} - \frac{1}{2} h_{\alpha\beta} \chi^{\alpha} \nabla_t \chi^{\beta} \\
    & +\frac{1}{4}G_{\mu\nu\alpha\beta}\lambda^{\mu}\lambda^{\nu}\chi^{\alpha}\chi^{\beta}
    - \frac{1}{3!}\nabla_{[\mu} c_{\nu\rho\sigma]}\lambda^{\mu}\lambda^{\nu}\lambda^{\rho}\lambda^{\sigma}\\
    & + \frac{1}{2}m_{\mu\alpha\beta}\Dot{q}^{\mu}\chi^{\alpha}\chi^{\beta} + m_{\mu\alpha\beta}\lambda^{\mu}y^{\alpha}\chi^{\beta} 
    -\frac{1}{2} \nabla_{[\mu} m_{\nu]\alpha\beta} \lambda^{\mu}\lambda^{\nu}\chi^{\alpha}\chi^{\beta} \\
    & + n_{\mu\nu\alpha} \Dot{q}^{\mu} \lambda^{\nu} \chi^{\alpha} 
    - \frac{1}{2} n_{\mu\nu\alpha} \lambda^{\mu} \lambda^{\nu} y^{\alpha} 
    - \frac{1}{2} \nabla_{[\mu} n_{\nu\rho]\alpha}\lambda^{\mu}\lambda^{\nu}\lambda^{\rho}\chi^{\alpha}  \\
    & - \frac{1}{2} l_{\alpha\beta\gamma}\chi^{\alpha}\chi^{\beta} y^{\gamma} 
    - \frac{1}{3!} \nabla_{\mu} l_{\alpha\beta\gamma} \lambda^{\mu}\chi^{\alpha}\chi^{\beta}\chi^{\gamma}\\
    &+ f_{\mu\alpha}\Dot{q}^{\mu}y^{\alpha} 
    + f_{\mu\alpha}\Dot{\lambda}^{\mu} \chi^{\alpha} 
    + \nabla_{\mu} f_{\nu\alpha} \lambda^{\mu}\Dot{q}^{\nu} \chi^{\alpha} \\
    & + A_{\mu}\Dot{q}^{\mu} -  \nabla_{[\mu} A_{\nu]} \lambda^{\mu}\lambda^{\nu} 
    - ms_{\alpha} y^{\alpha} - m \nabla_{\mu} s_{\alpha} \lambda^{\mu}\chi^{\alpha}
    \smash[t]{\biggr]}
  \end{split}
\end{equation}
where $G_{\mu\nu\alpha\beta}$ is the curvature of the vector bundle
connection $B_{\mu\phantom{\alpha}\beta}^{\phantom{\mu}\alpha}$,
\begin{equation}
  G^{\phantom{\mu\nu}\alpha}_{\mu\nu\phantom{\alpha}\beta}
  =\partial_{\mu}B_{\nu\phantom{\alpha}\beta}^{\phantom{\nu}\alpha}
  - \partial_{\nu}B_{\mu\phantom{\alpha}\beta}^{\phantom{\mu}\alpha}
  +B^{\phantom{\mu}\alpha}_{\mu\phantom{\alpha}\gamma}B^{\phantom{\nu}\gamma}_{\nu\phantom{\gamma}\beta}
  -B^{\phantom{\nu}\alpha}_{\nu\phantom{\alpha}\gamma}B^{\phantom{\mu}\gamma}_{\mu\phantom{\gamma}\beta}
\end{equation}
and $\nabla^{(+)}$ is the covariant derivative including the torsion,
\begin{equation}
  \nabla_t^{(+)}\lambda^{\mu} = \nabla_t\lambda^{\mu} - \frac{1}{2} c^{\mu}_{\phantom{\mu}\nu\rho}\Dot{q}^{\nu}\lambda^{\rho}
\end{equation}
The $N=1$ supersymmetry transformations of the component fields are
\begin{gather}
  \delta_\zeta q^{\mu} = \zeta \lambda^{\mu} \\
  \delta_\zeta \lambda^{\mu} = -\zeta \Dot{q}^{\mu} \\
   \delta_\zeta \chi^{\alpha} = \zeta (y^{\alpha} - \lambda^{\mu}B^{\phantom{\mu}\alpha}_{\mu\phantom{\alpha}\beta}\chi^{\beta})\\
  \delta_\zeta y^{\alpha} = -\zeta(\nabla_t
  \chi^{\alpha}+\lambda^{\mu}B^{\phantom{\mu}\alpha}_{\mu\phantom{\alpha}\beta}y^{\beta})
  +\frac{1}{2}\zeta\lambda^{\mu}\lambda^{\nu}\chi^{\beta} G^{\phantom{\mu\nu}\alpha}_{\mu\nu\phantom{\alpha}\beta}
\end{gather}

\section{Supergravity in one dimension}
\label{sec:local}
To construct the $N=1$ supergravity action, we gauge the rigid
supersymmetry by promoting the supersymmetry parameter $\zeta$ to a
local parameter, $\zeta=\zeta(t)$. In general this will destroy
invariance of (\ref{eq:rigidactioncom}) under supersymmetry, because we will get terms proportional
to $\Dot{\zeta}$. Therefore it is necessary to introduce gauge fields 
whose transformations will cancel with the $\Dot{\zeta}$ terms arising
from varying (\ref{eq:rigidactioncom}).

The method to find how these fields appear in the action, and their
transformations, is the Noether technique. We illustrate this
technique on a simple Lagrangian in flat space,
\begin{equation}
  \mathcal{L}_0=\frac{1}{2}\eta_{\mu\nu}\Dot{q}^{\mu}\Dot{q}^{\nu} + \frac{1}{2}\eta_{\mu\nu}\lambda^{\mu}\Dot{\lambda}^{\nu}
\end{equation}
Taking the supersymmetry transformation, with $\zeta=\zeta(t)$,
\begin{equation}
\label{eq:noetherexample}
  \delta_\zeta\mathcal{L}_0=\Dot{\zeta}\eta_{\mu\nu}\lambda^{\mu}\Dot{q}^{\nu}
\end{equation}
up to surface terms, which vanish in the action. To cancel
(\ref{eq:noetherexample}), consider the Lagrangian to first order in a
parameter $g$,
\begin{equation}
  \mathcal{L}_1=\mathcal{L}_0+g\psi\eta_{\mu\nu}\lambda^{\mu}\Dot{q}^{\nu}
\end{equation}
where $\psi$ is a gauge field with
$\delta_\zeta\psi=-g^{-1}\Dot{\zeta}$. Then the variation of the
new Lagrangian vanishes to zeroth order in $g$.

Continuing this process for all orders of $g$, we find
\begin{equation}
  \mathcal{L}=\frac{1}{2}e^{-1}\eta_{\mu\nu}\Dot{q}^{\mu}\Dot{q}^{\nu} +
    \frac{1}{2}\eta_{\mu\nu}\lambda^{\mu}\Dot{\lambda}^{\nu}
    +ge^{-1}\psi\eta_{\mu\nu}\Dot{q}^{\mu}\lambda^{\nu}
\end{equation}
It proves necessary to introduce a second gauge field $e$ which
transforms under local supersymmetry as $\delta_\zeta
e=2\zeta\psi$. It is an einbein which is the gauge field associated
with the diffeomorphisms of the worldline.

It is also necessary to modify
the transformation for $\lambda^{\mu}$ to
\begin{equation}
  \delta_\zeta \lambda^{\mu} = -e^{-1}\zeta(\Dot{q}^{\mu}+g\psi\lambda^{\mu})
\end{equation}
Applying the Noether method to (\ref{eq:rigidactioncom}), the $N=1$
supergravity action is
\begin{equation}
  \label{eq:locsc}
  \begin{split}
    S=\int\mspace{-5mu} \mathrm{d}t
    \biggl[ 
    & \frac{1}{2} e^{-1} g_{\mu\nu} (\Dot{q}^{\mu} + \psi\lambda^{\mu}) (\Dot{q}^{\nu} + \psi\lambda^{\nu})
    + \frac{1}{2} g_{\mu\nu} \lambda^{\mu} \nabla_t \lambda^{\nu} \\
    & - \frac{1}{2}eh_{\alpha\beta}y^{\alpha}y^{\beta} - \frac{1}{2} h_{\alpha\beta} \chi^{\alpha} \nabla_t \chi^{\beta} \\
    & +\frac{1}{4}e G_{\mu\nu\alpha\beta}\lambda^{\mu}\lambda^{\nu}\chi^{\alpha}\chi^{\beta}\\
    & + \frac{1}{2} c_{\mu\nu\rho}(\Dot{q}^{\mu}+\frac{2}{3}\psi\lambda^{\mu})\lambda^{\nu}\lambda^{\rho}
    - \frac{1}{3!}e\nabla_{[\mu} c_{\nu\rho\sigma]}\lambda^{\mu}\lambda^{\nu}\lambda^{\rho}\lambda^{\sigma} \\
    & + \frac{1}{2}m_{\mu\alpha\beta}\Dot{q}^{\mu}\chi^{\alpha}\chi^{\beta} + em_{\mu\alpha\beta}\lambda^{\mu}y^{\alpha}\chi^{\beta}
    -\frac{1}{2} e\nabla_{[\mu} m_{\nu]\alpha\beta} \lambda^{\mu}\lambda^{\nu}\chi^{\alpha}\chi^{\beta} \\
    & + n_{\mu\nu\alpha} (\Dot{q}^{\mu}+\frac{1}{2}\psi\lambda^{\mu})\lambda^{\nu} \chi^{\alpha}  
    - \frac{1}{2} e n_{\mu\nu\alpha} \lambda^{\mu} \lambda^{\nu} y^{\alpha}     
    - \frac{1}{2} e\nabla_{[\mu} n_{\nu\rho]\alpha}\lambda^{\mu}\lambda^{\nu}\lambda^{\rho}\chi^{\alpha}  \\
    & - \frac{1}{2} el_{\alpha\beta\gamma}\chi^{\alpha}\chi^{\beta} y^{\gamma} 
    - \frac{1}{3!} \psi l_{\alpha\beta\gamma}\chi^{\alpha}\chi^{\beta}\chi^{\gamma} 
    - \frac{1}{3!} e\nabla_{\mu} l_{\alpha\beta\gamma} \lambda^{\mu}\chi^{\alpha}\chi^{\beta}\chi^{\gamma}\\
    &+ f_{\mu\alpha}(\Dot{q}^{\mu}+\psi\lambda^{\mu})y^{\alpha} 
    + f_{\mu\alpha}\Dot{\lambda}^{\mu} \chi^{\alpha} 
    + \nabla_{\mu} f_{\nu\alpha} \lambda^{\mu}(\Dot{q}^{\nu}+\psi\lambda^{\nu}) \chi^{\alpha} \\
    & + A_{\mu}\Dot{q}^{\mu} - e \nabla_{[\mu} A_{\nu]} \lambda^{\mu}\lambda^{\nu} \\
    & - ms_{\alpha}(ey^{\alpha} + \psi\chi^{\alpha}) - em \nabla_{\mu} s_{\alpha} \lambda^{\mu}\chi^{\alpha}
    \smash[t]{\biggr]}
  \end{split}
\end{equation}
We observe that the Lagrangian is not obtained by the simple minimal
coupling rule $\Dot{q}^{\mu} \rightarrow \Dot{q}^{\mu} + \psi\lambda^{\mu}$.

The geometric interpretation of the couplings with the above action is
manifest. The above action is similar to that of models with rigid
supersymmetry in~\cite{Coles:1990hr}. After various redefinitions of
the couplings and the fields, one can recover the action constructed
in~\cite{vanHolten:1995qt} in the special case where
$f_{\mu\alpha}=0$. However, if $f_{\mu\alpha}\neq 0$
then~(\ref{eq:locsc}) is more general.

The $N=1$ supersymmetry transformations for the components fields become
\begin{gather}
  \label{eq:sgtra}
  \delta_\zeta q^{\mu} = \zeta\lambda^{\mu} \\ 
  \delta_\zeta \lambda^{\mu} = -\zeta e^{-1}(\Dot{q}^{\mu}+\psi\lambda^{\mu}) \\
   \delta_\zeta \chi^{\alpha} = \zeta (y^{\alpha} - \lambda^{\mu}B^{\phantom{\mu}\alpha}_{\mu\phantom{\alpha}\beta}\chi^{\beta})\\
  \delta_\zeta y^{\alpha} = -e^{-1}\zeta(\nabla_t
  \chi^{\alpha}+\psi y^{\alpha}) -\zeta\lambda^{\mu}B^{\phantom{\mu}\alpha}_{\mu\phantom{\alpha}\beta}y^{\beta}
  +\frac{1}{2}\zeta\lambda^{\mu}\lambda^{\nu}\chi^{\beta} G^{\phantom{\mu\nu}\alpha}_{\mu\nu\phantom{\alpha}\beta}
\end{gather}
The einbein $e$ and gravitino $\psi$ transform as
\begin{align}
  \delta_\zeta e &= 2\zeta\psi, & \delta_\zeta\psi &= -\Dot{\zeta}
\end{align}
Checking the algebra of the new transformations~(\ref{eq:sgtra}), 
\begin{align}
  [\delta_\zeta, \delta_\eta]q^{\mu} &= 2e^{-1}\zeta\eta (\Dot{q}^{\mu}+\psi\lambda^{\mu})
  & [\delta_\zeta, \delta_\eta]\lambda^{\mu} &= 2e^{-1}\zeta\eta (\Dot{\lambda}^{\mu} -e^{-1} \psi\Dot{q})\\
  &= 2e^{-1} ( \delta^{(H)}_{\zeta\eta} + \delta_{\zeta\eta\psi}) q^{\mu}  
  & &= 2e^{-1} ( \delta^{(H)}_{\zeta\eta} + \delta_{\zeta\eta\psi}) \lambda^{\mu} 
\end{align}
from which we obtain
\begin{equation}
  [ \delta_\zeta, \delta_\eta ] = 2e^{-1} ( \delta^{(H)}_{\zeta\eta} + \delta_{\zeta\eta\psi})
\end{equation}
and the same is true on the components $\chi^{\alpha},y^{\alpha}$. 

For invariance of (\ref{eq:locsc}) under worldline diffeomorphisms,
we need to specify the action of $\delta^{(H)}_{\epsilon}$ on the
einbein and the gravitino,
\begin{align}
  \delta_\epsilon^{(H)} e&=\partial_t(\epsilon e)
  & \delta_\epsilon^{(H)} \psi &= \partial_t(\epsilon\psi)
\end{align}
and those for $q^{\mu},\lambda^{\mu},\chi^{\alpha}$ and $y^{\alpha}$ are unchanged.

\section{Hamiltonian Analysis}
\label{sec:constraints}
To investigate the Hamiltonian dynamics of the system described by
(\ref{eq:elimaction}), we follow the Dirac-Bergman
procedure~\cite{Sundermeyer:1982gv} to analyse the constraints. This
will be important when we come to quantise the system in the next
section.

At this stage we introduce vielbeins $\mathrm{e}_{\mu}^{\phantom{\mu}i}$,
$\mathrm{f}_{\alpha}^{\phantom{\alpha}a}$ so that
$g_{\mu\nu}=\eta_{ij}\mathrm{e}_{\mu}^{\phantom{\mu}i}\mathrm{e}_{\nu}^{\phantom{\nu}j}$
and
$h_{\alpha\beta}=\eta_{ab}\mathrm{f}_{\alpha}^{\phantom{\alpha}a}\mathrm{f}_{\beta}^{\phantom{\beta}b}$
where $\eta_{ij}$ and $\eta_{ab}$ are the flat metrics on the
manifold and vector bundle respectively. We will use latin letters for
vielbein indices and greek letters otherwise. We take
\begin{align}
  \lambda^i &= \mathrm{e}_{\mu}^{\phantom{\mu}i}\lambda^{\mu},
  &\chi^a &= \mathrm{f}_{\alpha}^{\phantom{\alpha}a}\chi^{\alpha}
\end{align}
as our new fermion fields. This ensures that in the next section, the
Dirac brackets, hence commutation relations, between $p$ and the
fermions are zero. It is also necessary to set the Yukawa coupling
$f_{\mu\alpha}=0$.

Adopting this notation, and eliminating the auxiliary field
$y^{\alpha}$ from~(\ref{eq:rigidactioncom}) using its equation of
motion, gives the action
\begin{equation}
\label{eq:elimaction}
  \begin{split}
    S=\int\mspace{-5mu} \mathrm{d}t
    \biggl[ 
    & \frac{1}{2} g_{\mu\nu} e^{-1}(\Dot{q}^{\mu}+\psi
    \mathrm{e}^{\mu}_{\phantom{\mu}i}\lambda^{i})(\Dot{q}^{\nu}+\psi \mathrm{e}^{\nu}_{\phantom{\nu}j}\lambda^{j})
    + \frac{1}{2} \eta_{ij} \lambda^{i} \nabla_t \lambda^{j} 
    - \frac{1}{2} \eta_{ab} \chi^{a} \nabla_t \chi^{b} \\
    & +\frac{1}{2}eh_{ab}Y^{a}Y^{b} +\frac{1}{4}eG_{ijab}\lambda^{i}\lambda^{j}\chi^{a}\chi^{b}\\
    & + \frac{1}{2} c_{ijk}(\mathrm{e}_{\mu}^{\phantom{\mu}i}\Dot{q}^{\mu}+\frac{2}{3}\psi\lambda^i)\lambda^j\lambda^k
    - \frac{1}{3!}e \mathrm{e}^{\mu}_{\phantom{\mu}[i|}\nabla_{\mu} c_{|jkl]}\lambda^{i}\lambda^{j}\lambda^{k}\lambda^{l} \\
    & + \frac{1}{2}m_{\mu ab}\Dot{q}^{\mu}\chi^{a}\chi^{b} 
    -\frac{1}{2}e \mathrm{e}^{\mu}_{\phantom{\mu}[i|}\nabla_{\mu} m_{|j]ab} \lambda^{i}\lambda^{j}\chi^{a}\chi^{b} \\
    & + n_{ija} (\mathrm{e}_{\mu}^{\phantom{\mu}i}\Dot{q}^{\mu}+\frac{1}{2}\psi\lambda^{i}) \lambda^{j} \chi^{a} 
    - \frac{1}{2} e\mathrm{e}^{\mu}_{\phantom{\mu}[i|}\nabla_{\mu} n_{|jk]a}\lambda^{i}\lambda^{j}\lambda^{k}\chi^{a}  \\
    & -\frac{1}{3!}\psi l_{abc}\chi^{a}\chi^{b}\chi^{c}
    - \frac{1}{3!} e\mathrm{e}^{\mu}_{\phantom{\mu}i}\nabla_{\mu} l_{abc} \lambda^{i}\chi^{a}\chi^{b}\chi^{c}\\
    & + A_{\mu}\Dot{q}^{\mu} - e\mathrm{e}^{\mu}_{\phantom{\mu}[i|}\nabla_{\mu} A_{|j]} \lambda^{i}\lambda^{j} 
    - \psi ms_{a}\chi^{a}
    - em \mathrm{e}^{\mu}_{\phantom{\mu}i}\nabla_{\mu} s_{a} \lambda^{i}\chi^{a}
    \smash[t]{\biggr]}
  \end{split}
\end{equation}
where for convenience we define
\begin{equation}
\label{eq:defofya}
  Y_{a}=m_{iab}\lambda^{i}\chi^{b} -\frac{1}{2}n_{ija}\lambda^{i}\lambda^{j}
  -\frac{1}{2}l_{abc}\chi^{b}\chi^{c}  -ms_{a} 
\end{equation}
and the two covariant derivatives are with respect to the spin
connections $\omega_{\mu\phantom{k}l}^{\phantom{\mu}k}$ on the manifold
and $\Omega_{\mu\phantom{a}b}^{\phantom{\mu}a}$ on the vector bundle,
\begin{gather}
  \nabla_t \lambda^i = \partial_t \lambda^i + \Dot{q}^{\mu}\omega_{\mu\phantom{i}k}^{\phantom{\mu}i}\lambda^k\\
  \nabla_t \chi^a = \partial_t \chi^a + \Dot{q}^{\mu}\Omega_{\mu\phantom{a}b}^{\phantom{\mu}a}\chi^b
\end{gather}

The canonical momenta for $\lambda^i$ and $q^i$ are
\begin{gather}
  \pi_i = -\frac{1}{2}\eta_{ij}\lambda^j \\
  \begin{split}
    p_{\mu}  ={} &  e^{-1} g_{\mu\nu} (\Dot{q}^{\nu}+\psi \mathrm{e}^{\nu}_{\phantom{\nu}i}\lambda^i) 
    + \frac{1}{2}\omega_{\mu jk}\lambda^j\lambda^k
    -\frac{1}{2}\Omega_{\mu ab}\chi^a\chi^b\\
    &+ \frac{1}{2} \mathrm{e}_{\mu}^{\phantom{\mu}i} c_{ijk} \lambda^j\lambda^k 
    + \frac{1}{2}  m_{\mu ab}\chi^a\chi^b
    + \mathrm{e}_{\mu}^{\phantom{\mu}i} n_{ija}\lambda^j\chi^a 
    + A_{\mu} 
  \end{split}
\end{gather}
respectively. Similarly the canonical momenta for  $\chi^a$, $e$ and $\psi$ are
\begin{align}
  \pi_{\chi a} &= \frac{1}{2}\eta_{ab}\chi^b ,
  & \pi_e &= 0 ,
  & \pi_\psi &= 0
\end{align}
Clearly the system is constrained, as would be expected. The explicit
constraints are
\begin{align}
\label{eq:pc}
  \phi_i &= \pi_i+\frac{1}{2}\eta_{ij}\lambda^j\approx 0 
& \phi_{\chi a} &= \pi_{\chi a} - \frac{1}{2}\eta_{ab}\chi^b \approx 0 \\
  \phi_e &= \pi_e \approx 0
& \phi_\psi &= \pi_\psi \approx 0
\end{align}
where the $\approx$ denotes weak equality, in other words 
equality up to linear combinations of the other constraints.

The constrained Hamiltonian can then be found to be
\begin{equation}
  \begin{split}
    \mathcal{H}_c ={} & 
    \frac{1}{2}e \eta_{ij} P^i P^j
    - \psi \lambda^i  (P_i + \frac{1}{3}c_{ijk}\lambda^j\lambda^k 
    + \frac{1}{2}n_{ija}\lambda^j\chi^a)\\
    &- \frac{1}{2} e \eta_{ab} Y^a Y^b 
     - \frac{1}{4}eG_{ijab}\lambda^i\lambda^j\chi^a\chi^b\\
    &+ \frac{1}{3!} e \mathrm{e}^{\mu}_{\phantom{\mu}[i|}\nabla_{\mu}c_{|jkl]}\lambda^i\lambda^j\lambda^k\lambda^l 
    + \frac{1}{2} e \mathrm{e}^{\mu}_{\phantom{\mu}[i|}\nabla_{\mu} m_{|j]ab} \lambda^i\lambda^j\chi^a\chi^b\\
    &+ \frac{1}{2} e \mathrm{e}^{\mu}_{\phantom{\mu}[i|}\nabla_{\mu}n_{|jk]a}\lambda^i\lambda^j\lambda^k\chi^a
    + \frac{1}{3!} e \mathrm{e}^{\mu}_{\phantom{\mu}i}\nabla_{\mu} l_{abc} \lambda^i\chi^a\chi^b\chi^c \\
    &+ \frac{1}{3!} \psi l_{abc}\chi^a\chi^b\chi^c
    + e \mathrm{e}^{\mu}_{\phantom{\mu}[i|}\nabla_{\mu} A_{|j]} \lambda^i\lambda^j \\
    &+ m \psi s_a\chi^a 
    + e \mathrm{e}^{\mu}_{\phantom{\mu}i}\nabla_{\mu} m s_a \lambda^i\chi^a 
  \end{split}
\end{equation}  
where $Y^a$ was defined in (\ref{eq:defofya}) and 
\begin{equation}
  \begin{split}
    P_i ={} &  \mathrm{e}^{\mu}_{\phantom{\mu}i} p_{\mu} - \frac{1}{2}\omega_{ijk}\lambda^j\lambda^k
    +\frac{1}{2}\Omega_{iab}\chi^a\chi^b\\
    &- \frac{1}{2} c_{ijk} \lambda^j\lambda^k 
    - \frac{1}{2} m_{iab}\chi^a\chi^b
    - n_{ija}\lambda^j\chi^a 
    - A_i 
  \end{split}
\end{equation}
The primary Hamiltonian is defined to be
\begin{equation}
  \mathcal{H}_p = \mathcal{H}_c + \phi_i u^i +
                  \phi_{\chi a} u_{\chi}^{\phantom{\chi}a}
                  + \phi_e u_e + \phi_\psi u_\psi
\end{equation}
where the $u$ are all Lagrange multipliers (and $u^i$,
$u_{\chi}^{\phantom{\chi}a}$, $u_{\psi}$ are anticommuting).

We require that the constraints~(\ref{eq:pc}) hold for all time, 
\begin{align}
\label{eq:sccon}
  \Dot{\phi}_i &= \{\phi_i,\mathcal{H}_p\} \approx 0
&  \Dot{\phi}_{\chi a} &= \{\phi_{\chi a},\mathcal{H}_p\} \approx 0\\
\label{eq:fccon}
  \Dot{\phi}_e &= \{\phi_e,\mathcal{H}_p\} \approx 0
&  \Dot{\phi}_\psi &= \{\phi_\psi,\mathcal{H}_p\} \approx 0
\end{align}
Each condition either determines a multiplier or leads to a new
constraint. We assume the canonical Poisson brackets
\begin{align}
  \{q^\mu, p_\nu\} &= \delta^\mu_{\phantom{\mu}\nu} \\
  \{\chi^a,\pi_{\chi b} \} &= -\delta^a_{\phantom{a}b} \\
  \{\lambda^i, \pi_j \} &= -\delta^i_{\phantom{i}j} \\
  \{e,\pi_e\} &=1 \\
  \{\psi,\pi_\psi\} &= -1
\end{align}
Imposing (\ref{eq:fccon}) requires the secondary constraints
\begin{gather}
\label{eq:Hconstraint}
  \begin{split}
    \varphi_e ={} & \frac{1}{2}\eta_{ij}P^iP^j - \frac{1}{2}\eta_{ab}Y^aY^b
     -\frac{1}{4}G_{ijab}\lambda^i\lambda^j\chi^a\chi^b \\
    & + \frac{1}{3!} \mathrm{e}^{\mu}_{\phantom{\mu}[i|}\nabla_{\mu} c_{|jkl]}\lambda^i\lambda^j\lambda^k\lambda^l 
     + \mathrm{e}^{\mu}_{\phantom{\mu}[i|}\nabla_{\mu} A_{|j]}\lambda^i\lambda^j\\
   &  + m \mathrm{e}^{\mu}_{\phantom{\mu}i}\nabla_{\mu} s_a\lambda^i\chi^a 
    + \frac{1}{2} \mathrm{e}^{\mu}_{\phantom{\mu}[i|}\nabla_{\mu} m_{|j]ab}\lambda^i\lambda^j\chi^a\chi^b\\
   &  + \frac{1}{2} \mathrm{e}^{\mu}_{\phantom{\mu}[i|}\nabla_{\mu} n_{|jk]a} \lambda^i\lambda^j\lambda^k\chi^a
     + \frac{1}{3!} \mathrm{e}^{\mu}_{\phantom{\mu}i}\nabla_{\mu} l_{abc}\lambda^i\chi^a\chi^b\chi^c
  \end{split}\\
\label{eq:Qconstraint}
    \varphi_\psi =  
    \lambda^i (P_i+\frac{1}{3}c_{ijk}\lambda^j\lambda^k +\frac{1}{2}n_{ija}\lambda^j\chi^a)
    - \frac{1}{3!}l_{abc}\chi^a\chi^b\chi^c
    - ms_a \chi^a 
\end{gather}
In fact these are the charges of $H,Q$ respectively. It can be checked
that both of these are conserved over time,
\begin{align}
  \{ \varphi_e, \mathcal{H}_p \} &\approx 0 &   \{ \varphi_\psi, \mathcal{H}_p \} &\approx 0 
\end{align}
so they give rise to no new constraints.

The remaining conditions (\ref{eq:sccon}) determine
\begin{align}
\label{eq:multipliers}
  u^i &= e \eta^{ij}\{ \varphi_e,\phi_j \} + \psi \eta^{ij} \{ \varphi_\psi, \phi_j \} \\
  u_{\chi}^{\phantom{\chi}a} &= 
  -e \eta^{ab} \{ \varphi_e,\phi_{\chi b} \} 
  - \psi \eta^{ab}\{ \varphi_\psi, \phi_{\chi b} \}
\end{align}

Observe that the constrained Hamiltonian $\mathcal{H}_c$ can be
written in terms of the secondary constraints
\begin{equation}
\label{eq:cansch}
  \mathcal{H}_c = \varphi_e e + \varphi_\psi\psi
\end{equation}
Essentially this is because $\varphi_e=\{\pi_e,\mathcal{H}_c\}$ and
$\varphi_\psi=\{\pi_\psi,\mathcal{H}_c\}$ and the Hamiltonian
$\mathcal{H}_c$ is linear in the gauge fields $e$ and $\psi$.

\section{Quantisation}
\label{sec:quantise}
We observe that the constraints $\phi_i$ and $\phi_{\chi a}$ are both
second class, whereas
\begin{gather}
  \phi_e \\
  \phi_\psi \\
  \varphi'_e = \varphi_e + \eta^{ij}\{ \varphi_e,\phi_i \}\phi_j
             - \eta^{ab}\{ \varphi_e,\phi_{\chi a}\}\phi_{\chi b} \\
  \varphi'_\psi = \varphi_\psi + \eta^{ij}\{ \varphi_\psi,\phi_i \}\phi_j
             - \eta^{ab}\{ \varphi_\psi,\phi_{\chi a}\}\phi_{\chi b} 
\end{gather}
are all first class. Defining the Dirac bracket as
\begin{equation}
\label{eq:defdirac}
  \{ A, B \}_D = \{A,B\} + \eta^{ij}\{ A,\phi_i\}\{\phi_j, B\} 
  - \eta^{ab}\{ A,\phi_{\chi a}\}\{\phi_{\chi b}, B\}
\end{equation}
then Poisson brackets between the primed constraints are weakly equal
to Dirac brackets between the original unprimed constraints. 

The extra terms on the right of~(\ref{eq:defdirac}) give rise to new
relations
\begin{align}
\label{eq:diracbrackets}
  \{\lambda^i, \lambda^j \}_D &= \eta^{ij} \\
  \{\chi^a,\chi^b\}_D &=-\eta^{ab}
\end{align}

We can check that the $N=1$ supersymmetry algebra is still obeyed,
\begin{equation}
  \{\varphi'_\psi,\varphi'_\psi\} \approx \{\varphi_\psi,\varphi_\psi\}_D = 2\varphi_e \approx 0 
\end{equation}
and that all the brackets between secondary constraints vanish,
\begin{align}
  \{\phi_e,\varphi'_e\} &\approx \{\phi_e,\varphi_e\}_D = 0
&  \{\phi_{\psi},\varphi'_\psi\} &\approx \{\phi_{\psi},\varphi_\psi\}_D = 0 \\
  \{\varphi'_e,\varphi'_e\} &\approx \{\varphi_e,\varphi_e\}_D = 0 
&  \{\varphi'_e,\varphi'_\psi\} &\approx \{\varphi_e,\varphi_\psi\}_D = 0 
\end{align}

The Hamiltonian can be written in terms of these first
class constraints. From~(\ref{eq:multipliers}) and~(\ref{eq:cansch}),
\begin{equation}
  \mathcal{H}_p = \varphi_e e + \varphi_\psi \psi + \phi_eu_e + \phi_\psi u_\psi
\end{equation}
so it is vanishing weakly, as is expected for a gravitational system.

In Dirac's process of quantisation, fields become operators acting on some
Hilbert space and second class constraints are imposed as operator conditions
on the states. First class constraints generate unphysical degrees of
freedom, so it is necessary to fix a gauge,
\begin{align}
  e &= 1 & \psi &= 0
\end{align}

Moving over to the quantised system, Dirac brackets become
(anti) commutation relations. One realisation of this algebra is
using the standard Clifford algebra generators
\begin{align}
  \{ \gamma^i, \gamma^j \} &= 2\eta^{ij} & \{ \gamma^a, \gamma^b \} &= 2\eta^{ab}
\end{align}
If the dimension of the manifold is even, for example $d=4$, then we
have an element of the algebra, $\gamma^{d+1}$, satisfying
$(\gamma^{d+1})^2=-1$ and $\{ \gamma^{d+1}, \gamma^i \}=0$ so the
following realisation exists,
\begin{align}
  \Hat{\lambda}^i &= \frac{1}{\sqrt{2}} \gamma^i \otimes 1
  & \Hat{\chi}^a &= \frac{1}{\sqrt{2}} \gamma^{d+1} \otimes \gamma^a
\end{align}
The $\gamma^{d+1}$ ensures that $\Hat{\lambda}^i$ and $\Hat{\chi}^a$
anticommute.  

If the manifold $\mathcal{M}$ is not even dimensional then the
following realisation may be used,
\begin{align}
    \Hat{\lambda}^i &= \frac{1}{2} \gamma^i \otimes 1 \otimes \sigma_1
  & \Hat{\chi}^a &= \frac{1}{2} 1 \otimes \gamma^a \otimes \sigma_2
\end{align}
where $\sigma_i$ are the Pauli spin matrices which satisfy
$\{\sigma_i,\sigma_j\}=2\delta_{ij}$. For the $N$-extended case, see
for example~\cite{Gates:1995ch}.

In the following we will assume that the dimension of $\mathcal{M}$ is
even and use the first realisation given. Then the second class
constraints are imposed as conditions on physical states,
\begin{align}
  \Hat{Q}|\mathrm{phys}\rangle&=0
  & \Hat{H}|\mathrm{phys}\rangle&=0
\end{align}
where 
\begin{multline}
  \label{eq:dirac}
     \Hat{Q}=-\frac{1}{\sqrt{2}}(\gamma^i \otimes 1)\mathrm{e}^{\mu}_{\phantom{\mu}i}\frac{\partial}{\partial q^{\mu}}
     - \frac{1}{4\sqrt{2}}\omega_{ijk}(\gamma^{i} \gamma^{jk} \otimes 1)
    - \frac{1}{4\sqrt{2}}\Omega_{iab} (\gamma^i \otimes \gamma^{ab})\\
    + \frac{1}{12\sqrt{2}} c_{ijk} (\gamma^{ijk} \otimes 1)
    - \frac{1}{4\sqrt{2}} m_{iab}(\gamma^i \otimes \gamma^{ab})
    - \frac{1}{4\sqrt{2}}n_{ija}(\gamma^{ij}\gamma^{d+1} \otimes \gamma^a)\\
    - \frac{1}{\sqrt{2}}A_i (\gamma^i \otimes 1)
    + \frac{1}{12\sqrt{2}}l_{abc}(\gamma^{d+1}\otimes\gamma^{abc})
    - \frac{1}{\sqrt{2}}ms_a (\gamma^{d+1}\otimes\gamma^a)
\end{multline}
\begin{multline}
  \label{eq:kleingordon}
    \Hat{H}=\frac{1}{2}\eta_{ij}\Hat{P}^i\Hat{P}^j - \frac{1}{2}\eta_{ab}\Hat{Y}^a\Hat{Y}^b
    +\frac{1}{16}R_{ijkl}(\gamma^{ij}\gamma^{kl}\otimes 1) 
    +\frac{1}{16}G_{ijab}(\gamma^{ij}\otimes\gamma^{ab})\\
     + \frac{1}{24} \mathrm{e}^{\mu}_{\phantom{\mu}[i|} \nabla_{\mu} c_{|jkl]}(\gamma^{ijkl}\otimes 1)
    - \frac{1}{8} \mathrm{e}^{\mu}_{\phantom{\mu}[i|} \nabla_{\mu} m_{|j]ab}(\gamma^{ij}\otimes\gamma^{ab})\\
     + \frac{1}{8} \mathrm{e}^{\mu}_{\phantom{\mu}[i|}\nabla_{\mu} n_{|jk]a} (\gamma^{ijk}\gamma^{d+1}\otimes\gamma^a)
     - \frac{1}{24} \mathrm{e}^{\mu}_{\phantom{\mu}i} \nabla_{\mu} l_{abc}(\gamma^i\gamma^{d+1}\otimes\gamma^{abc})\\
     + \frac{1}{2}\mathrm{e}^{\mu}_{\phantom{\mu}[i|}\nabla_{\mu} A_{|j]}(\gamma^{ij}\otimes 1)
     + \frac{1}{2}m \mathrm{e}^{\mu}_{\phantom{\mu}i} \nabla_{\mu} s_a(\gamma^i\gamma^{d+1}\otimes \gamma^a) \\
     - \omega_{\phantom{mi}m}^{mi}\mathrm{e}^\mu_{\phantom{\mu}i}\frac{\partial}{\partial
     q^\mu} 
    -\frac{1}{8}\omega^{mi}_{\phantom{mi}m}\omega_{ikl}(\gamma^{kl}\otimes 1)\\
    - \frac{1}{2}\Gamma^{\mu}_{\phantom{\mu}\nu\rho}g^{\nu\rho}\frac{\partial}{\partial q^\mu}
    + \frac{1}{24}c^{ijk}c_{ijk} - \frac{1}{24} l^{abc} l_{abc}\\
    -\frac{1}{16}c_{ijk}\omega^{mi}_{\phantom{mi}m}(\gamma^{jk}\otimes 1)
    -\frac{1}{8}n_{ija}\omega^{mi}_{\phantom{mi}m}(\gamma^j\gamma^{d+1}\otimes\gamma^a)
    +\frac{1}{8}m_{iab}\omega^{mi}_{\phantom{mi}m}(1\otimes \gamma^{ab})    
\end{multline}
The new terms which appear here include a Riemann tensor which vanishes
classically because it is contracted with four fermions, and several
terms involving a trace of the connection which arise because of the
way we have chosen to order the equation.

In~(\ref{eq:dirac}) and~(\ref{eq:kleingordon}) we have defined as usual
\begin{equation}
  \begin{split}
      \Hat{P}_i ={} &
    -\mathrm{e}^{\mu}_{\phantom{\mu}i}\frac{\partial}{\partial q^{\mu}} 
    - \frac{1}{4}\omega_{ijk}(\gamma^{jk}\otimes1)
    -\frac{1}{4}\Omega_{iab}(1\otimes\gamma^{ab})\\
    &- \frac{1}{4} c_{ijk} (\gamma^{jk}\otimes 1)
    + \frac{1}{4} m_{iab}(1\otimes\gamma^{ab})
    - \frac{1}{2}n_{ija}(\gamma^j\gamma^{d+1}\otimes\gamma^a)
    - A_i 
  \end{split}
\end{equation}
and
\begin{equation}
    \Hat{Y}_a={} 
    \frac{1}{2}m_{iab}(\gamma^i\gamma^{d+1}\otimes\gamma^b)
  -\frac{1}{4}n_{ija}(\gamma^{ij}\otimes 1)
  +\frac{1}{4}l_{abc}(1\otimes\gamma^{ab}) 
  -ms_a(1\otimes 1)
\end{equation}
It can be checked that 
\begin{align}
  \{\Hat{Q},\Hat{Q}\}=2\Hat{H}
\end{align}
the familiar result that the square of the Dirac operator gives the
Klein Gordon equation.

Finally we note that it is not always the case that the system will
have physical states. This is because manifolds exist for which the
Dirac-like operators do not have zero modes. However it is expected
that most models will have a physical Hilbert space which is non-empty.

\providecommand{\href}[2]{#2}\begingroup\raggedright\endgroup

\end{document}